# Understanding Enhanced Mechanical Stability of DNA in the Presence of Intercalated Anticancer Drug: Implications for DNA Associated Processes

Anil Kumar Sahoo,[1] Biman Bagchi,[2] and Prabal K. Maiti[1,*]

[1] Center for Condensed Matter Theory, Department of Physics, Indian Institute of Science, Bangalore-560012, India

[2] Solid State and Structural Chemistry Unit, Indian Institute of Science, Bangalore-560012, India

* To whom correspondence to be addressed. Email: maiti@iisc.ac.in

## ABSTRACT

Most of the anticancer drugs bind to double-stranded DNA (dsDNA) by intercalative-binding mode. Although experimental studies have become available recently, a molecular-level understanding of the interactions between the drug and dsDNA that lead to the stability of the intercalated drug is lacking. Of particular interest are the modifications of the mechanical properties of dsDNA observed in experiments. The latter could affect many biological functions, such as DNA transcription and replication. Here we probe, via all-atom molecular dynamics (MD) simulations, change in the mechanical properties of intercalated drug–DNA complexes for two intercalators, daunomycin and ethidium. We find that, upon drug intercalation, stretch modulus of DNA increases significantly, whereas its persistence length and bending modulus decrease. Steered MD simulations reveal that it requires higher forces to stretch the intercalated dsDNA complexes than the normal dsDNA. Adopting various pulling protocols to study force-induced DNA melting, we find that the dissociation of dsDNA becomes difficult in the presence of intercalators. The results obtained here provide a plausible mechanism of function of the anticancer drugs—i.e., via altering the mechanical properties of DNA. We also discuss long-time consequences of using these drugs, which require further in vivo investigations.

## I. INTRODUCTION

The interaction of a small drug molecule with dsDNA is a subject of great current interest. A small-molecule drug can interact with a dsDNA either specifically or non-specifically. There are three specific DNA-binding modes—namely, minor groove, major groove and intercalative bindings—along with the non-specific electrostatic binding mode.[1] In intercalative binding, a drug (or part of it) inserts between two consecutive base-pairs of a dsDNA [see Figs. 1(b) and 1(c)]. All intercalative drugs have common design principle: the drug consists of phenyl rings arranged in a plane and a positively charged side chain. The phenyl rings intercalate into DNA and are stabilized by the π-stacking interactions with the DNA bases, while the side chain interacts electrostatically with the negatively charged DNA backbone. Enormous interest in understanding the process of drug-intercalation and its effect on DNA mechanics exists because most of the anticancer drugs are known to intercalate into DNA. DNA-intercalators are believed to hinder DNA replication and transcription,[2-4] eventually leading to cell death—thus acting as anticancer drugs. However, serious side effects of the intercalating anticancer drugs have been observed due to low selectivity towards cancer cells and DNA sequences. It is important to obtain and understand the mechanical and biophysical properties of drug–DNA complexes for the design and optimization of new anticancer drugs, as these drugs mainly target DNA.[5,6]

Over the last two decades, significant progress has been made to experimentally study various biophysical processes, such as protein folding/unfolding[7,8] and DNA mechanics,[9,10] and enormous insight has been gained about the intermediate structures, free-energy landscape, pathways and kinetics of conformational transitions—thanks to the advancement in single-molecule force spectroscopy (SMFS) techniques.[11] More recently, SMFS techniques have been used to study in details the kinetics of intercalation and the mechanism of DNA and small-molecules (or proteins) interactions along with the possible biological implications.[1,12-14] Intercalators have also been used as fluorescent probes to unravel structural details about various overstretched conformations of DNA.[15-17] The overstretching transition of dsDNA[15,18] is known to shifts towards higher forces at lower intercalator concentration; however at a high enough concentration of intercalator, the overstretched state vanishes completely.[3,19,20] Some experimental works have demonstrated that intercalators increase DNA melting force.[19,21,22] Recent SMFS experiments suggest that intercalators increase the stretch (or Young's) modulus of DNA,[23,24] whereas the bending rigidity and persistence length of dsDNA remain unaffected upon drug intercalation.[25,26] Despite these experimental studies, there is still controversy about the modifications in elastic properties of dsDNA in the presence intercalators.[13] These experimental results have also not been studied by simulations, except in references.[22,23,27] In particular, little seems to be known about the details of the effects of small-molecule binding on the functions of dsDNA, such as its dissociation mechanics.

Molecular dynamics (MD) simulation studies have provided valuable insights into various aspects of the intercalation process.[28,29] Particularly, for an anticancer drug daunomycin,[30] the pathways, the free energy landscapes and the thermodynamics of intercalation have been studied in great details, revealing the formation of a DNA minor-groove-bound intermediate state before a fully intercalated state is reached.[31,32] For another simple intercalator such as ethidium (also frequently used as a fluorescent tag), the changes in the dsDNA structure and dynamics upon intercalation, the end-state thermodynamics of intercalation and dsDNA dissociation in presence of the intercalator have been investigated.[22,33] These studies along with other experimental studies revealed that intercalators distort dsDNA structure locally, such as an increase in the base-pair rise along with specific changes in the twist and roll parameters, and that intercalation is enthalpically driven, while groove-binding is entropically driven.[34] However, despite having important implications for biological functions and pharmacology, systematic determination and quantitative understanding of the changes in elastic properties of DNA, its dissociation mechanics and pathways, and its structure in presence of intercalators are far from complete.

In this study, we consider two DNA intercalators, daunomycin and ethidium. We have performed all-atom equilibrium and nonequilibrium MD simulations in explicit solvent to unravel the microscopic mechanics of the intercalated drug–DNA complexes as well as of the normal B-DNA, for comparison. The rest of the manuscript is organized as follows. First, molecular modeling approach and simulation protocols are described in the Methods section. Then, we discuss and compare the equilibrium simulations results of the intercalated drug–DNA complexes and the B-DNA, followed by those from nonequilibrium simulations. We conclude this article with a plausible mechanism of action of the anticancer drugs, possible implications of using it and some outlooks.

## II. METHODS

### A. Model Building and Force Field Parameters

The crystal structure for the complex, daunomycin drug intercalated to dsDNA, was obtained from the protein data bank (PDB ID: 1D11). Following the procedure by Mukherjee et al.,[31,32] we built a dsDNA dodecamer of sequence (GCGCA**CG**TGCGC)$_2$; the intercalation site is present between the central C6 and G7 bases (denoted in bold characters in the above sequence). A segment of 5′-ACG-3′ from the crystal structure was taken that contains a single daunomycin molecule between C and G bases. To build the complete structure, this segment was further extended by adding in both sides appropriate numbers of base-pairs in the canonical B-form by using the nucleic acid builder (NAB) tool.[35] A similar procedure was followed to build dsDNA–ethidium complex, except the central part, i.e., ethidium intercalated between the two GC base pairs was taken from the available crystal structure.[36] The same DNA sequence as the DNA–daunomycin complex was chosen for better comparison. For a control simulation, a B-DNA, i.e., dsDNA dodecamer of the above sequence in the canonical B-form was built by using the NAB tool.[35]

The ff14SB[37] force field parameters with parmbsc0[38] correction were used for DNA. The molecular models for daunomycin and ethidium were developed following Merz/Kollman method.[39] First, their geometry optimizations and the ESP charge calculations were done through the Gaussian package[40] using a 6-31G* basis set. Then, the ANTECHAMBER[41] module of AMBER was used to obtain the RESP charges on each atom and the GAFF[42] atom types. Each complex was solvated with TIP3P[43] water model using the *xleap* module of the AMBER17 tools[44] in a large enough rectangular box guaranteeing no image interactions. The box dimensions of the solvated B-DNA, dsDNA–daunomycin and dsDNA–ethidium complexes are 7.4×7.4×7.4 nm$^3$, 8.0×8.0×8.0 nm$^3$ and 8.1×8.3×8.3 nm$^3$, respectively. Appropriate numbers of monovalent ions (22 $Na^+$ and 1 $Cl^-$) were added in the simulation box ensuring the charge neutrality of the unit shell. Joung/Cheatham[45] ion parameters were used to describe the interaction between ions, water and DNA.

**B. Equilibrium MD Simulation Protocol**

Periodic boundary condition (PBC) was used for all simulations. Bonds involving hydrogen atoms were constrained using the SHAKE[46] algorithm that allows using a larger time step of 2 fs for the integration of the equations of motion. The temperature of the system was maintained using a Langevin thermostat[47] with the collision frequency of 5.0 ps$^{-1}$. The Berendsen weak coupling method[48] was used to maintain a pressure of 1 atm with a pressure relaxation time constant of 2.0 ps. The particle mesh Ewald (PME)[49] sum was used to compute the long-range electrostatic interactions with a real-space cutoff of 10 Å. The van der Waals (vdW) and direct electrostatic interactions were truncated at the cutoff. The direct sum non-bonded list was extended to cutoff + "nonbond skin" (10 + 2 Å).

The solvated system with harmonic restraints (force constant of 500 kcal mol$^{-1}$ Å$^{-2}$) on the position of each solute atom was first subjected to 1000 steps of steepest descent energy minimization, followed by 2000 steps of conjugate gradient minimization to remove bad contacts present in the initially built systems. Restraints on the solute atoms were sequentially decreased to zero during further 4000 steps of energy minimization. The energy minimized system was then slowly heated from 10 to 300 K in many steps during the first 21 ps of MD simulation. During this dynamic, the solute particles were restrained to their initial positions using harmonic restraints with a force constant of 20 kcal mol$^{-1}$ Å$^{-2}$. The initial restrained heating and the next first 2 ns equilibration simulations were performed in the NPT ensemble to reach the proper density. Subsequently, 200 ns simulation was run in the NVT ensemble. Simulations were performed using the PMEMD module of the AMBER14 package.[50]

**C. Steered Molecular Dynamics (SMD) Simulation Protocol**

The equilibrated structure of dsDNA–daunomycin (or dsDNA–ethidium) complex from AMBER simulation was taken as the starting structure, re-solvated in a rectangular box (the size varies depending on the pulling protocol as described in Fig. 2) and added counter ions for charge neutralization. This solvated system was subjected to energy minimization and heating protocols as described in the previous section, prior to equilibration in NPT ensemble. The temperature of the systems was controlled at 300 K by using a Langevin thermostat[47] with a collision frequency of 5.0 ps$^{-1}$. 1 atm pressure was maintained using Nose−Hoover Langevin barostat (see section 7.5.2 of NAMD user's guide, http://www.ks.uiuc.edu/Research/namd/2.10b1/ug.pdf) with piston period of 0.2 ps and damping time constant of 0.05 ps. PME[49] was used to compute the long-range electrostatic interactions with a real-space cutoff of 12 Å. The vdW interaction was truncated at a cutoff of 12 Å by using a switch function (switchdist = 10 Å). PBC and a time step of 2 fs (enabled via SHAKE)[46] were used. All equilibrium and subsequent SMD simulations were performed using the NAMD-2.10 package.[51]

For the SMD simulations depending on the pulling protocol as described in Fig. 2, some atoms of the dsDNA were fixed, and some other atoms of the dsDNA were pulled. In a SMD simulation,[52] the atom being pulled (called SMD atom) is connected via a spring (here, force constant $k = 10$ kcal mol$^{-1}$ Å$^{-2}$) to a virtual atom, which is moved with a constant velocity $V$ in a direction $\vec{n}$ that depends on the pulling protocol (see the direction of the arrow in Fig. 2). The effective SMD potential, $U$, is given by $U(\vec{r}) = \frac{1}{2}k[Vt - (\vec{r} - \vec{r_0}) \cdot \vec{n}]^2$, where $\vec{r}$ and $\vec{r_0}$ are the positions of the atom being pulled at time $t$ and at the initial time, respectively. And the force, $\vec{F}$, on the pulled atom is calculated by:

$$\vec{F} = -\vec{\nabla}U = k[Vt - (\vec{r} - \vec{r_0}) \cdot \vec{n}]\vec{n}. \tag{1}$$

In our previous work,[53] we used a similar SMD protocol to study protein–protein binding in the presence and absence of dendrimer. All the SMD simulations performed in this work add to a total simulated time of ~5.5 µs for systems containing ≥ 60661 atoms. At a pulling velocity $V = 1$ Å/ns, individual SMD simulations for the SE, OS3, OS5, EU and MU protocols (see Fig. 2) were performed up to 40 ns, 100 ns, 100 ns, 150 ns and 60 ns, respectively. For the SE pulling protocol, individual SMD simulations at $V = 0.2$ Å/ns and $V = 0.05$ Å/ns were performed up to 200 ns and 800 ns, respectively. For the MU protocol, each SMD simulation at $V = 0.2$ Å/ns was performed up to 270 ns.

### D. Data Analysis

All the analyses were done by using in-house codes and the AMBER17 tools.[44] Images were rendered using the VMD software.[54]

Curves+ software[55] was used to calculate the helix axis and various inter base pair parameters of dsDNA, which include the three distances (shift, slide, and rise) and three angles (twist, roll, and tilt) that completely describe successive base pair planes. The distance, $l$, between centers of successive base pairs in a step is given by $l = \sqrt{(shift)^2 + (slide)^2 + (rise)^2}$.[56] These data were used to calculate the end-to-end distance, contour length, and the local tangent vectors.

The extension for each pulling protocol (see Fig. 2) is defined as $Vt$, i.e., is the distance traveled by the virtual atom which is connected via a spring to the SMD atom.

The total number of hydrogen bonds, $N_{HB}$, is calculated by summing the hydrogen bonds present for each DNA base pair. A hydrogen bond is counted if the distance between a donor atom (Dn) and an acceptor atom (Ac) is ≤ 3 Å and the Dn–H⋯Ac angle is ≤ 135°.

The average cumulative work done $\langle W \rangle$ for taking the DNA from its equilibrium configuration to a deformed state is given by:

$$\langle W \rangle = \langle \int_0^t dt\, \vec{V} \cdot \vec{F} \rangle, \tag{2}$$

where $\langle \cdot \rangle$ represents the average over four different simulation runs, and all other symbols are the same as those in Eq. (1). The estimated error in the work at a given extension is the standard deviation of four different simulation runs at a pulling velocity $V = 1$ Å/ns, as shown for the different cases in Figs. 3–8.

## III. RESULTS

### A. DNA mechanics and the effect of drug intercalation on various structural and elastic properties of DNA from equilibrium simulations

The mechanical properties of dsDNA can be inferred by studying its various structural properties. Two of the important phenomenological parameters that describe DNA elasticity are the stretch and bending moduli. Here, we calculate these two quantities. We also calculate structural properties such as end-to-end distance, contour length, and persistence length of the bare B-DNA [see Fig. 1(a)] and DNA in the presence of an intercalated daunomycin [see Fig. 1(b)] or ethidium [see Fig. 1(c)]. All the results are summarized in Table I.

End-to-end distance, $L_d$, of DNA is defined as the distance between the centers-of-mass of the two terminal base pairs of the dsDNA. We see ~4 Å increase in $L_d$ for both the intercalated DNA complexes compared to that for the bare B-DNA. Contour length, $L_0$, is calculated by summing over the distance ($l_i$) between each base pair step and it is given by $L_0 = \sum_{i=1}^{N-1} l_i$, where $N$ is the total number of base pairs. Like $L_d$, we observe that $L_0$ increases for the drug intercalated DNA complexes. These increases in $L_d$ and $L_0$ upon drug-intercalation is due to an increase in the rise between the two base pairs that accommodate the intercalated drug, as shown in Fig. S1(a) in the supplementary material (SM).

Stretch modulus, $\gamma$, is obtained from the contour length distribution as described below, following our earlier works.[57,58] In equilibrium, the instantaneous contour length, $L$, of DNA fluctuates around its mean value $L_0$. Within the elastic rod model, the restoring force, $F$, generated due to the instantaneous fluctuation in the contour length, $L - L_0$, is proportional to $L - L_0$, and it is given by $F = -\gamma(L - L_0)/L_0$. The energy function, $E(L)$, due to $F$ is obtained by integrating $F$ w.r.t. $L$ and is given by $E(L) = \gamma(L - L_0)^2/2L_0$. Then, the probability distribution for the contour length, $P(L)$, is obtained by putting the energy term, $E(L)$, in the Boltzmann's formula and normalizing it. $P(L)$ is a Gaussian distribution and is given by:

$$P(L) = \sqrt{\frac{\beta\gamma}{2\pi L_0}} \exp\left[-\frac{\beta\gamma L_0}{2}\left(\frac{L}{L_0} - 1\right)^2\right] \tag{3}$$

$$\Rightarrow \ln P(L) = -\frac{\beta\gamma L_0}{2}\left(\frac{L}{L_0} - 1\right)^2 + \frac{1}{2}\ln\frac{\beta\gamma}{2\pi L_0}, \tag{4}$$

where $\beta = 1/k_BT$; $k_B$ is the Boltzmann constant, and $T$ is the temperature. $P(L)$ obtained from our simulation (data points) fits well with Eq. (3) (lines) for all the complexes, as shown in Fig. 1(d) [left panel]. γ is then obtained from the slope in Fig. 1(d) [right] by fitting the simulation data using Eq. (4). The average stretch moduli of all the complexes and the associated errors are provided in Table I. For the bare B-DNA, we find that the stretch modulus γ is 1167 pN, which is in quantitative agreement with the values estimated in previous experiments as well as our earlier simulations.[57-60] γ of the DNA–daunomycin complex is 1535 pN and is 32% higher than that of the bare DNA. For the DNA–ethidium complex, we find that the stretch modulus of DNA is even higher (2119 pN), which is 81% higher than the value obtained for the bare DNA. The increase in γ of DNA upon a drug intercalation might be due to the enhancement in the π-π stacking interaction among the base pairs near to the intercalation site, because of the interaction between the bases and the phenyl rings of the intercalator.

Bending modulus, κ, (and persistence length, $l_P = \beta\kappa$) can be determined from the bending angle distribution as follows. To calculate the bending angle (θ) of DNA, we define two tangent vectors by joining the centers-of-mass of two consecutive terminal base pairs for either ends of dsDNA, and θ is obtained by taking the cosine inverse of the dot product of these two vectors. The probability distribution for the bending angle, $P(\theta)$, obtained from the simulation for each complex is plotted in Fig. 1(e) [left]. $P(\theta)$ for small fluctuations in θ can be approximated as Gaussian in nature,[61] and is given by:[57,58]

$$P(\theta) = \sqrt{\frac{\beta\kappa}{2\pi L_0}} \exp\left[-\frac{\beta\kappa}{2L_0}\theta^2\right] \tag{5}$$

$$\Rightarrow \ln P(\theta) = -\frac{l_P}{L_0}(1 - \cos\theta) + \frac{1}{2}\ln\frac{\beta\kappa}{2\pi L_0}. \tag{6}$$

Persistence length, $l_P$, and bending modulus ($\kappa = l_P k_B T$) are obtained from the slope of $(1 - \cos\theta)$ versus $\ln P(\theta)$ plot in Fig. 1(e) [right] by a linear fit of simulation data to Eq (6). The average bending moduli and persistence lengths of all the complexes and the associated errors are given in Table I. In contrast to the behavior observed for stretch modulus, we find that $l_P$ (and hence κ) of dsDNA decreases by 16% and 41% in the presence of daunomycin and ethidium, respectively. This decrease in $l_P$ (and κ) might be due to the local unwinding of the DNA double helix structure near to the intercalation site [see Fig. S1(b) in the SM], which facilitates large fluctuation in the bending angle $\theta$.

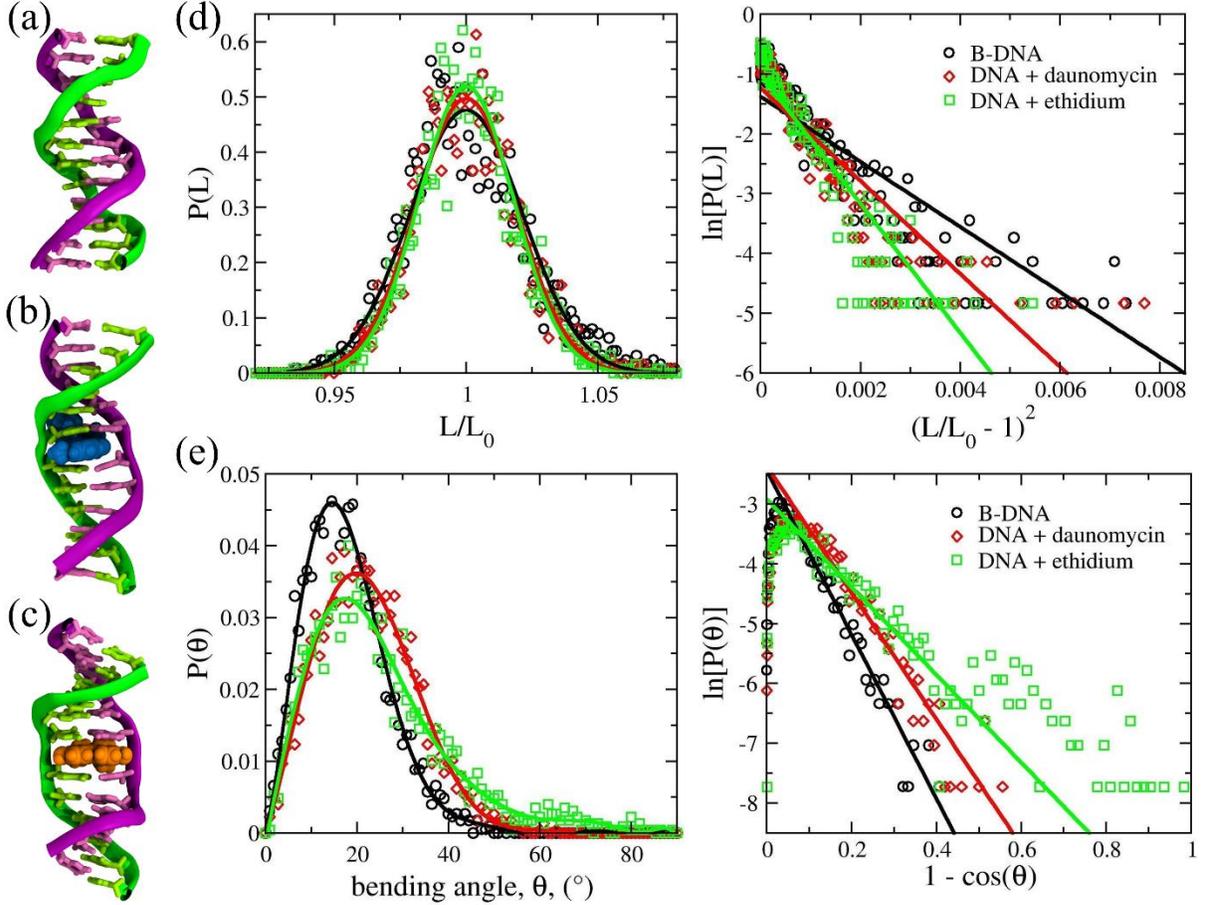

FIG. 1. Comparison of structure and elastic properties of the B-DNA and the intercalated drug–DNA complexes from equilibrium simulations. (a) Equilibrated structure of the B-DNA: One strand of the dsDNA is represented in purple color, while another strand is depicted in green color. (b) Equilibrated structure of the DNA–daunomycin complex: The drug daunomycin (color: blue) intercalated between the middle two base-pairs of the dsDNA is shown in VDW representation. (c) Equilibrated structure of the DNA–ethidium complex, where the drug ethidium is depicted in orange color. Water molecules and ions are not shown in (a), (b) and (c) for clarity. Stretch modulus, $\gamma$, and persistence length, $l_P$, (as well as bending rigidity $\kappa$) for each system are calculated from the normalized probability distributions of DNA contour length (d) and bending angle (e), respectively. See the text for details. The distributions for each system are obtained from the last 50 ns of 200 ns long simulation.

TABLE I. The structural and elastic properties of dsDNA in the presence and absence intercalated drugs calculated from equilibrium simulations data and their comparisons with experiments. The errors in $L_d$ and $L_0$ are the standard deviations of the respective data points obtained from the last 50 ns of 200 ns long simulation. The errors in $\gamma$, $\kappa$ and $l_P$ are the propagation errors in multiplications of $L_0$ with the respective slopes (see Fig. 1). The error in the slope is the standard

error of the estimate in linear regression. The experimental values for the bare DNA are taken from the study by Gross et al.[60]

| System | End-to-end distance, $L_d$, (Å) | Contour length, $L_0$, (Å) | Stretch modulus, $\gamma$, (pN) | Bending modulus, $\kappa$, (pN·nm$^2$) | Persistence length, $l_P$, (nm) |
|---|---|---|---|---|---|
| B-DNA | 35.38±1.07 | 38.63±0.88 | 1166.9±72.3 | 219.42±8.58 | 52.97±2.07 |
| DNA + daunomycin | 39.06±1.00 | 41.92±0.82 | 1534.6±88.5 | 183.58±8.23 | 44.32±1.99 |
| DNA + ethidium | 39.35±1.05 | 42.18±0.79 | 2118.6±102.3 | 128.19±7.32 | 30.97±1.77 |
| bare DNA (experiment) | — | 28500±60 | 1450±50 | 162±12 | 39±3 |

*The "local" effect of an intercalator on the structural and elastic properties of DNA*

As can be seen from Fig. S1 in the SM that an intercalator affects the DNA base-stacking interactions maximum up to 3 base pairs from the intercalation site. To understand what is the extent to which the intercalators locally modify various properties of DNA, we consider the middle 6 base-pairs of the DNA duplex (denoted hereafter as the half-length DNA). The structural and elastic properties of the half-length DNA are calculated by following the same methods as discussed for the full-length DNA. The specific details and analyses for the half-length DNA for all the three systems are presented in Fig. S2 in the SM. The structural and elastic parameters of the half-length DNA for all the three systems are provided in Table II. We find 54% and 86% increases in the stretch modulus of the half-length DNA in the presence of daunomycin and ethidium, respectively. In contrast, the persistence lengths (and hence the bending modulus) of the half-length DNA for the DNA–daunomycin and DNA–ethidium complexes decrease by 32% and 38%, respectively. These results are in accord with the results for the full-length DNA complexes. As expected, we find that the effect of an intercalator on the structural and mechanical properties of DNA is more pronounced in the vicinity of an intercalated drug.

TABLE II. The structural and elastic properties of the middle 6 base pairs of dsDNA in the presence and absence of intercalated drugs calculated from equilibrium simulations data. The errors in $L_d$ and $L_0$ are the standard deviations of the respective data points obtained from the last 50 ns of 200 ns long simulation. The errors in $\gamma$, $\kappa$ and $l_P$ are the propagation errors in multiplications of $L_0$ with the respective slopes (see Fig. S2 in the SM). The error in the slope is the standard error of the estimate in linear regression.

| System | End-to-end distance, $L_d$, (Å) | Contour length, $L_0$, (Å) | Stretching modulus, $\gamma$, (pN) | Bending modulus, $\kappa$, (pN·nm$^2$) | Persistence length, $l_P$, (nm) |
|---|---|---|---|---|---|
| B-DNA | 16.06±0.65 | 17.52±0.66 | 880.5±73.5 | 252.09±17.93 | 60.86±4.33 |
| DNA + daunomycin | 19.75±0.51 | 20.68±0.57 | 1359.9±108.0 | 171.16±9.22 | 41.32±2.23 |
| DNA + ethidium | 19.61±0.93 | 21.01±0.57 | 1639.8±110.4 | 155.99±8.44 | 37.66±2.04 |

**B. DNA mechanics and the effect of drug intercalation on it from SMD simulations**

Most of the biological processes in the cell operate at far-from-equilibrium, with the help of mechanical forces generated consuming chemical energy. Specifically, in DNA transcription and replication processes, a dsDNA must be partially melted/dissociated for read-out of the genetic codes. DNA intercalators might hamper these processes as intercalators modify the elastic properties of dsDNA, as shown in the previous section. Besides, nonlinear effects become prominent in presence of higher stretching forces and it is interesting to see how that affects the mechanics of the intercalated drug–DNA complexes. Motivated by these questions, we perform constant velocity pulling simulations for the bare dsDNA and the DNA intercalated complexes by using the SMD simulation protocol (described in the Methods section) for five different pulling geometries as depicted in Fig. 2. We discuss and compare below the results for the B-DNA with the DNA–daunomycin complex for each pulling geometry, followed by the results for the DNA–ethidium complex for two different pulling geometries.

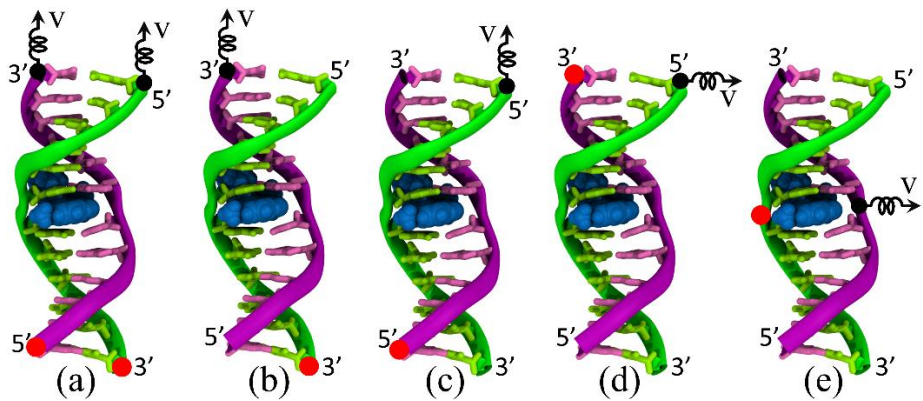

FIG. 2. Various constant velocity pulling simulation protocols for the daunomycin intercalated dsDNA dodecamer. (a) The same end (SE) pulling protocol: We fix one end of each DNA strand those are present on the same side and pull the opposite side ends of both strands along the long axis of the DNA. The representation and color coding for the DNA–daunomycin complex is the same as described in Fig. 1. The fixed atoms are shown as red circles. The SMD atoms, connected via a spring to a virtual atom that is being pulled with velocity $V$, are depicted as black circles. The arrow represents the direction of pulling. (b) The OS3 pulling protocol: We fix 3′-end of one strand and pull another strand's 3′-end along the long axis of DNA. (c) The OS5 pulling protocol: 5′-end of one strand is fixed, and another strand's 5′-end is pulled along the long axis of DNA. (d) The end-unzipping (EU) protocol: We fix 3′-end of one strand and pull 5′-end of another strand but in a direction perpendicular to the long axis of DNA. (e) The mid-unzipping (MU) protocol: The middle backbone phosphorus atom of one DNA strand is fixed, while the central backbone phosphorus atom of another strand is pulled in a direction perpendicular to the long axis of DNA. Note that while the base-pairs are sheared for the pulling protocols used in (a), (b) and (c), for the 'tear mode' protocols adopted in (d) and (e) the base-pairs are unzipped one by one.

*1. DNA stretching by the SE pulling protocol*

The pulling geometry for the SE pulling protocol is described in Fig. 2(a). Fig. 3 (left) shows the force–extension profile for the DNA–daunomycin complex. For comparison, we have also shown the force–extension profile for the bare B-DNA in the same plot. For the DNA–daunomycin complex, we observe that the stretching force $F$ [defined in Eq. (1)] increases linearly up to an extension of 10 Å (the first elastic regime), followed by a plateau region for 10–25 Å where $F$ merely increases (the overstretching transition), and a sharp increase in $F$ after 25 Å (the second elastic regime). This force–extension profile is qualitatively very similar to the bare B-DNA as shown in Fig. 3. Presence of such three different force regimes in the DNA force–extension behavior has also been found in earlier experimental[18] and simulation[62,63] studies. Note that $F$ for the intercalated complex is found to be more than that for the B-DNA over the entire range of extensions. In Fig. 3 (right), we show the instantaneous snapshots of the DNA–daunomycin complex at various extensions. We see that the DNA elongates while its two strands unwind up to an extension of 25 Å (see snapshots i–v in Fig. 3). Beyond that extension, the DNA backbones are stretched (see snapshots vi–viii in Fig. 3) giving rise to the sharp increase in $F$. The snapshots show that many of the DNA base pairs melt during this process. To quantify this, we have calculated the total number of hydrogen bonds, $N_{HB}$, as DNA is pulled (see the Methods for the definition). $N_{HB}$ versus extension plot shows that all the base pairs remain intact up to the extension of 15 Å; then the base pairs gradually melt (see Fig. 3). We find that at very high extension (40 Å), $N_{HB}$ for the DNA–daunomycin complex is more than that for the B-DNA—many base pairs present nearer to the intercalator remain intact. Surprisingly, daunomycin remains intercalated, even after significant deformations of dsDNA structure (see snapshot viii in Fig. 3). This is due to the strong π-π interaction between the phenyl rings of the intercalator and the enclosing DNA base pairs. We also calculate the average cumulative work done $\langle W \rangle$ [see Eq. (2)] for stretching dsDNA from its equilibrium configuration and find that $\langle W \rangle$ for the DNA–daunomycin complex lies above that for the B-DNA for all extensions, except the initial 10 Å elongations for which both curves overlap (Fig. 3). So, it is much harder to stretch the intercalated drug–DNA complex compared to the B-DNA. This observation is in accordance with our finding that the stretch modulus of dsDNA significantly increases upon drug intercalation (see Table I).

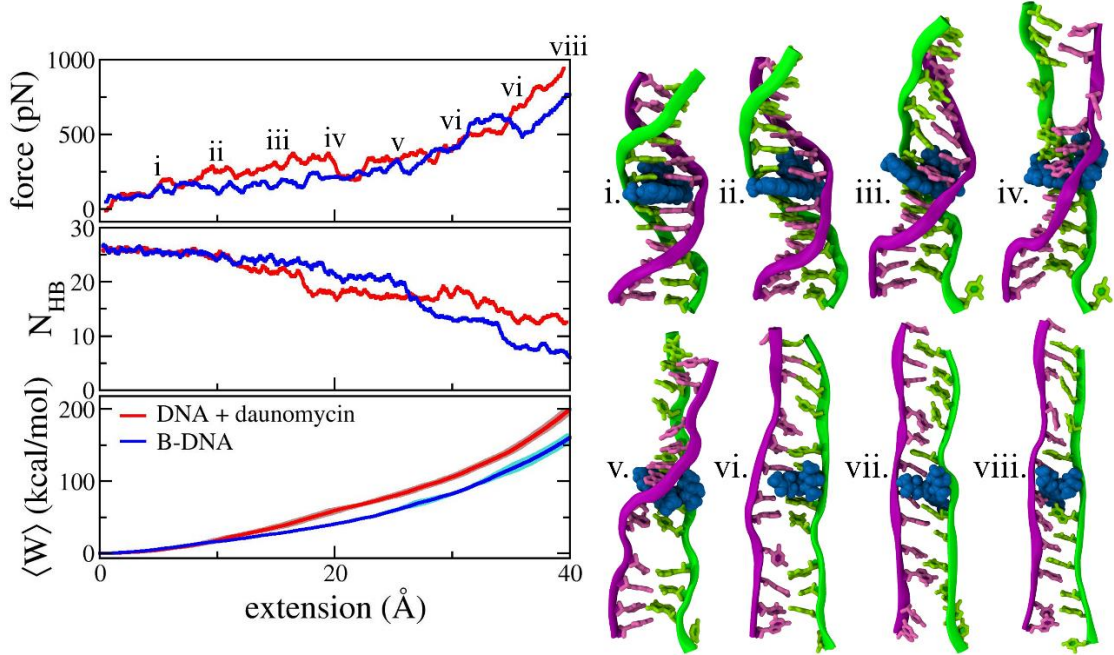

FIG. 3. SMD simulation results for the SE pulling protocol as described in Fig. 2(a) for $V = 1$ Å/ns. [left] From top to bottom: Stretching forces, the total number of hydrogen bonds ($N_{HB}$) involved in forming the DNA base pairs and the average cumulative work, $\langle W \rangle$, at various DNA-extensions for the DNA–daunomycin complex, and their comparisons with the B-DNA results. The associated error bars in the calculations of $\langle W \rangle$s for the DNA–daunomycin complex and the B-DNA are shown in brown and cyan colors, respectively. [right] Snapshots (i–viii) represent the DNA–daunomycin complex at various DNA-extensions as pointed in the force–extension plot. The representation and color coding are the same as described in Fig. 1.

## 2. DNA shearing by the OS3 pulling protocol

The OS3 pulling protocol is described in Fig. 2(b), and all the results are summarized in Fig. 4. We see from the force–extension plot that $F$ initially increases linearly up to 15 Å followed by a force-plateau region for 15–37 Å. Then $F$ again increases linearly for the extensions of 37–62 Å, and it finally decreases to zero with a sharp jump at 62 Å. From snapshots i–iii in Fig. 4 for the DNA–daunomycin complex at various extensions, we find that the dsDNA elongates while its two strands unwind up to the extension of 37 Å. After complete unwinding, the dsDNA elongates further where its backbones are stretched, and its base pairs are sheared. The dsDNA forms a ladder-like structure for this intermediate-range of extensions (see snapshots iv–v in Fig. 4), where most of its base pairs remain intact as evident from the $N_{HB}$ versus extension plot (Fig. 4). This is the "B-to-S" transition—a dsDNA present in the B-form when overstretched goes to an extended conformation called S-DNA[15,18]—that has been confirmed in both experiments[16,64,65] and simulations.[63,66] Here, we find that the intercalated drug–DNA complex adapts S-DNA like structure when pulled from the 3′-end of the DNA. Unlike the gradual decrease of $N_{HB}$ as for the SE pulling protocol, $N_{HB}$ decreases slightly from its initial value up to the extension of 62 Å. After that extension, $N_{HB}$ sharply drops down to zero representing melting of most of the base pairs (snapshots vi–vii in Fig. 4), which eventually leads to complete separation of the two DNA strands (snapshot viii in Fig. 4). So, the force peak observed ~62 Å is mainly due to the tension build-up in the hydrogen-bonded base pairs. The rupture force, i.e., the maximum force required to separate the two strands is 980 pN for the B-DNA, whereas it is 1200 pN for the DNA–daunomycin complex. We find that, for all extensions, the average cumulative work $\langle W \rangle$ for the intercalated complex is more than that for the B-DNA. $\langle W \rangle$ for the complete separation of the two DNA strands for the drug intercalated complex is 37 kcal/mol more than that of the bare B-DNA. So, it is difficult to melt a dsDNA in the presence of intercalators by shearing it. We also observe that the base pairs enclosing the intercalator break at the end (snapshots vii–viii in Fig. 4). This fact supports the finding that the strong π-π stacking interaction between the intercalator and the enclosing DNA base pairs is the cause for the enhanced stability of the intercalated complex, as discussed for the SE pulling protocol.

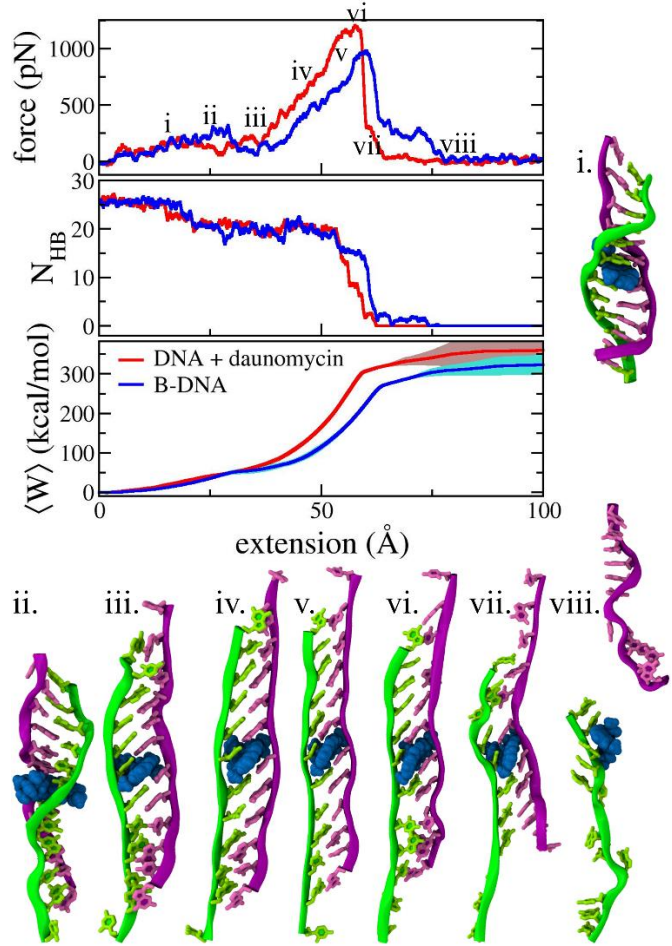

FIG. 4. SMD simulation results for the OS3 pulling protocol as described in Fig. 2(b) for $V = 1$ Å/ns. $F$, $N_{HB}$ and $\langle W \rangle$ at various extensions for the DNA–daunomycin complex and the bare B-DNA are plotted. The associated error bars in the calculations of $\langle W \rangle$s for the DNA–daunomycin complex and the B-DNA are shown in brown and cyan colors, respectively. Snapshots (i–viii) represent the DNA–daunomycin complex at various DNA-extensions as pointed in the force–extension plot. The representation and color-coding in each snapshot are the same as described in Fig. 1.

### *3. DNA shearing by the OS5 pulling protocol*

Here, we pull the DNA from its 5′-end. This pulling protocol is described in Fig. 2(c), and all the results are summarized in Fig. 5. As in the OS3 pulling protocol, we find similar force–extension behavior. The force initially increases linearly up to 15 Å. There is a force plateau region for extensions of 15–37 Å followed by a linear increase in $F$ up to ~60 Å; finally, $F$ drops down to zero suddenly. But, the rupture forces for the B-DNA (575 pN) and the DNA–daunomycin complex (840 pN) obtained in this protocol are much less than the corresponding values for the OS3 pulling protocol. This can be understood by looking at the $N_{HB}$ plot and the snapshots at various extensions (see Fig. 5). $N_{HB}$ only remain constant up to 20 Å, and it decreases smoothly to zero over the extensions range of 20–60 Å. So, the base pairs melt gradually as the DNA elongates. This contrasts with the trend observed for the OS3 pulling protocol, in which most base pairs remain intact up to 60 Å balancing the tension. We find, here, no observation of any ladder-like S-DNA structure for either the intercalated complex (see snapshots ii–vi in Fig. 5) or the B-DNA. As in earlier experimental[67] and simulation[63] studies of force-induced DNA melting, we also find no signature of any transformation from the B-DNA to the S-DNA conformation by pulling the B-DNA from its 5′-end, whereas such a "B-to-S" transition was found when pulled from its 3′-end. As for the OS3 pulling protocol, the base pairs near to the intercalator breaks at the end, resulting in separation of the two DNA strands (snapshots vi–viii in Fig. 5). The average cumulative work $\langle W \rangle$ for the complete separation of the two strands of the DNA–daunomycin complex is 46 kcal/mol more than that of the B-DNA. We find that, for this protocol also, intercalator make it harder to dissociate a dsDNA by shearing it.

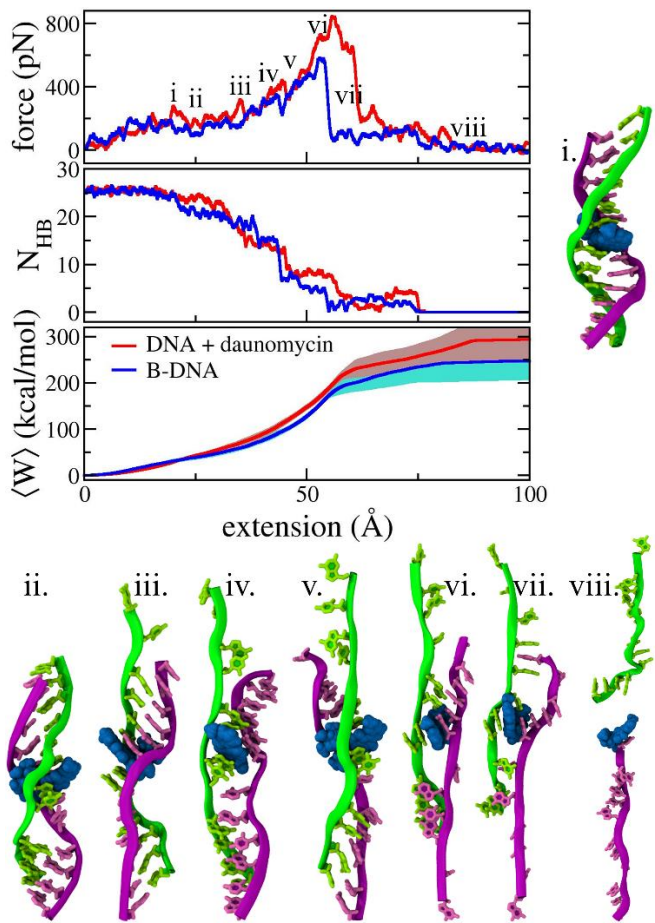

FIG. 5. SMD simulation results for the OS5 pulling protocol as described in Fig. 2(c) for $V = 1$ Å/ns. $F$, $N_{HB}$ and $\langle W \rangle$ at various extensions for the DNA–daunomycin complex and the bare B-DNA are plotted. The associated error bars in the calculations of $\langle W \rangle$s for the DNA–daunomycin complex and the B-DNA are shown in brown and cyan colors, respectively. Snapshots (i–viii) represent the DNA–daunomycin complex at various DNA-extensions as pointed in the force–

extension plot. The representation and color-coding in each snapshot are the same as described in Fig. 1.

## *4. DNA tearing by unzipping it from one end (EU)*

We unzip the dsDNA to get an idea about the length, i.e., the number of bases from the intercalation site up to which the base pairs are more stabilized due to the additional stacking interaction with the intercalator, apart from the usual intra-base-pair hydrogen bonds and base-base stacking interactions. The pulling geometry for the EU protocol is described in Fig. 2(d), and all the results are summarized in Fig. 6. The force–extension plot is very different from those discussed before, showing a sawtooth pattern. There are many force ($F$)-peaks of ~250 pN, each corresponding to the breaking of a base pair. This is clear from the stepwise decrease in the $N_{HB}$ as a function of the extension (see also the instantaneous snapshots i–vi in Fig. 6). Comparing the unzipping forces for the B-DNA and the DNA–daunomycin complex, we find that the $F$-peaks for the latter is higher than the former, except for the first three terminal-base-pairs. The $F$-peaks are similar for breaking the first three terminal-base-pairs of both the complexes. Also, from the average cumulative work $\langle W \rangle$ plot in Fig. 6, we see that $\langle W \rangle$ for the intercalated complex is more than that for the B-DNA only after extension of 37 Å; otherwise it is the same for both. We find that, for the complete separation of the two DNA strands, $\langle W \rangle$ for the intercalated complex is 44 kcal/mol more than that for the B-DNA. Thus, intercalator helps in stabilizing the base pairing and base-base stacking interactions strengths up to three bases far away from the intercalation site.

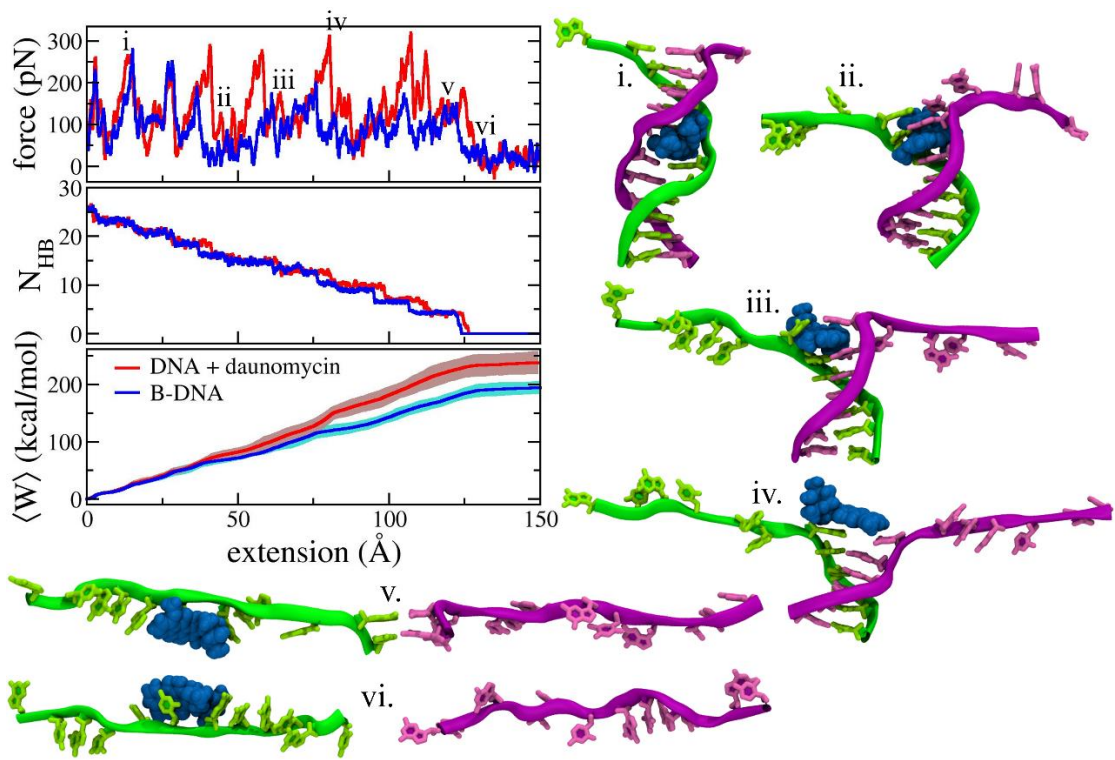

FIG. 6. SMD simulation results for the EU protocol as described in Fig. 2(d) for $V = 1$ Å/ns. $F$, $N_{HB}$ and $\langle W \rangle$ at various extensions for the DNA–daunomycin complex and the bare B-DNA are plotted. The associated error bars in the calculations of $\langle W \rangle$s for the DNA–daunomycin complex and the B-DNA are shown in brown and cyan colors, respectively. Snapshots (i–vi) represent the DNA–daunomycin complex at various DNA-extensions as pointed in the force–extension plot. The representation and color-coding in each snapshot are the same as described in Fig. 1.

## *5. DNA tearing by unzipping it from the middle (MU)*

The dsDNA structure and energetics in the vicinity of an intercalated drug are affected the most, as we have seen in the above discussion. To directly probe these, we pull the two DNA strands from its middle. The pulling geometry for the MU protocol is described in Fig. 2(e), and all the results are summarized in Fig. 7. The unzipping force, $F$, initially increases linearly with a peak at the extension of 5 Å. Then $F$ drop down and fluctuates between 100–300 pN. It finally becomes zero after the extension of 50 Å, representing the complete separation of the two DNA strands. The peak forces for the B-DNA and the DNA–daunomycin are 640 pN and 1000 pN, respectively. The force peak corresponds to the simultaneous breaking of the two middle base pairs (see snapshot i in Fig. 7). This is clear from the drop in $N_{HB}$ at an extension of 5 Å in Fig. 7. $N_{HB}$ then slowly decreases as the extension increases, with a final rapid drop to zero in between 45–50 Å. The force–extension plots for the B-DNA and the intercalated complex almost lie on top of each other for all extensions, except for 0–14 Å. This shows that once the nearby base pairs on either side of the intercalator are broken, the effect of the intercalator is minimal (see the snapshots in Fig. 7). The average cumulative work $\langle W \rangle$ for the complete separation of the two strands of the drug intercalated complex is 52 kcal/mol more than that for the bare B-DNA. Note that for this protocol $\langle W \rangle$s for the complete separation of the B-DNA and the intercalated complex are the lowest among the respective values for all the different pulling protocols.

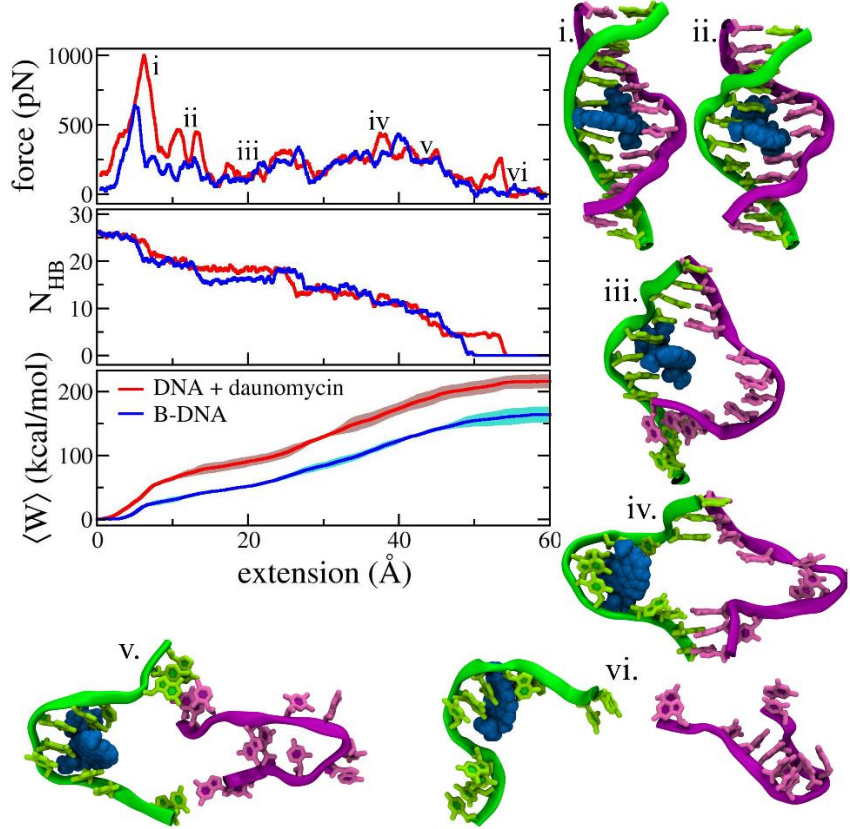

FIG. 7. SMD simulation results for the MU protocol as described in Fig. 2(e) for $V = 1$ Å/ns. $F$, $N_{HB}$ and $\langle W \rangle$ at various extensions for the DNA–daunomycin complex and the bare B-DNA are plotted. The associated error bars in the calculations of $\langle W \rangle$s for the DNA–daunomycin complex and the B-DNA are shown in brown and cyan colors, respectively. Snapshots (i–vi) represent the DNA–daunomycin complex at various DNA-extensions as pointed in the force–extension plot. The representation and color-coding in each snapshot are the same as described in Fig. 1.

## *6. DNA shearing and tearing in the presence of ethidium intercalator*

To check the effect of ethidium on DNA mechanics, we have also performed SMD simulations for two different pulling protocols: the OS3 pulling and the MU. We choose these two protocols as the cumulative works, $\langle W \rangle$s, for the complete separation of two strands of the B-DNA are found to be the maximum and minimum for the OS3 pulling and the MU protocols, respectively. The OS3 pulling and the MU protocols are described in Figs. 2(b) and 2(e), respectively. As shown in Fig. 8(a), we observe similar shearing force (and $N_{HB}$) versus extension behavior(s) for the OS3 pulling protocol, as in the case of DNA–daunomycin complex (see Fig. 4). The rupture force for the DNA–ethidium complex is 1240 pN. This is comparable to the rupture force for the DNA–daunomycin complex (1200 pN) but larger than that for the bare B-DNA (980 pN). The average cumulative work $\langle W \rangle$ for the complete separation of the two DNA strands of the DNA–ethidium complex is 345 kcal/mol. This work is more than the work required for the complete dissociation of the bare B-DNA (323 kcal/mol) but lower than the same for the DNA–daunomycin complex which is 360 kcal/mol. As for the B-DNA and the DNA–daunomycin complex in case of the OS3 pulling protocol, we also observe a ladder-like S-DNA conformation for the DNA–ethidium complex, which can be seen from the instantaneous snapshots iii–vi shown in Fig. S3 in the SM. For the MU protocol also, as for the DNA–daunomycin (Fig. 7), we observe similar tearing force (and $N_{HB}$) versus extension behavior(s) for the DNA–ethidium as shown in Fig. 8(b). The instantaneous snapshots at various extensions are provided in Fig. S4 in the SM. Although the peak force for the DNA–ethidium complex (850 pN) is more than that for the B-DNA (640 pN), it is much less than the peak force for the DNA–daunomycin complex (1000 pN). $\langle W \rangle$ for the complete separation of the two strands of the DNA–ethidium complex is 172 kcal/mol. When compared to $\langle W \rangle$ for the complete dissociation of the B-DNA, the total work done for the DNA–ethidium complex is only 8 kcal/mol more, while the same for the DNA–daunomycin complex is 52 kcal/mol higher. Therefore, the intercalation of ethidium strengthens the dsDNA complex as we observe from both the pulling protocols, though the effect is less compared to daunomycin intercalation.

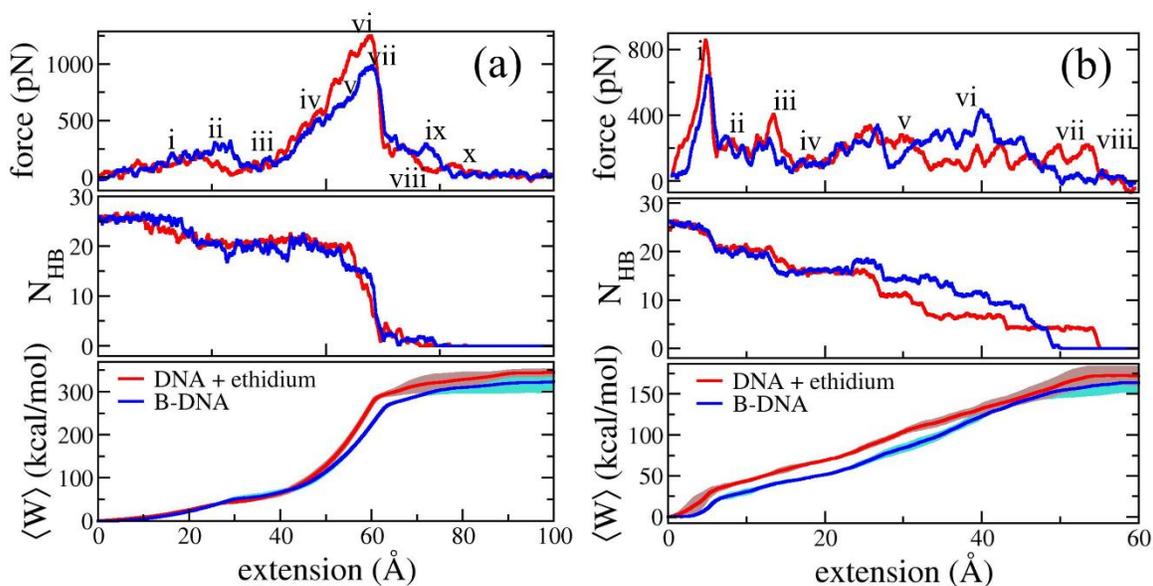

FIG. 8. SMD simulation results at $V = 1$ Å/ns for the DNA–ethidium complex [Fig. 1(c)], and their comparisons with the B-DNA results. From top to bottom: comparison of forces, $N_{HB}$s and $\langle W \rangle$s between the DNA–ethidium complex and the B-DNA for the OS3 pulling (a) and the MU (b) protocols. The OS3 pulling and the MU protocols are described in Figs. 2(b) and 2(e), respectively. The associated error bars in the calculations of $\langle W \rangle$s for the DNA–daunomycin complex and the B-DNA are shown in brown and cyan colors, respectively. The instantaneous snapshots at various extension as marked in the force–extension plots for the OS3 pulling and the MU protocols are shown in Figs. S3 and S4 in the SM, respectively.

## IV. DISCUSSION

The total work done, $\langle W \rangle$, for the B-DNA melting by the OS3 pulling, OS5 pulling, EU and MU protocols are 323, 247, 195 and 164 kcal/mol, respectively. $\langle W \rangle$ for melting the DNA–daunomycin complex by the OS3 pulling, OS5 pulling, EU and MU protocols, respectively are 360, 294, 239 and 216 kcal/mol, while the same for melting the DNA–ethidium complex by the OS3 pulling and MU protocols are 345 and 172 kcal/mol, respectively. We see that $\langle W \rangle$ very much depends on the pulling protocol. For dsDNA melting in case of each of the three complexes, $\langle W \rangle$ is the lowest for the MU protocol, whereas it is the highest for the OS3 pulling protocol. Note that if the dsDNA complex is separated in truly equilibrium manner, then the free energy or the reversible $\langle W \rangle$ for separating the two DNA strands should be equal for all the pulling protocols. This pulling protocol dependence of $\langle W \rangle$ can be rationalized by considering the additional contribution to the free energy due to the dissipative forces coming from the hydrogen-bonding friction.[68] Depending on the pulling protocol the intra-base-pair hydrogen bonds are broken apart by either shearing or tearing the dsDNA base pairs, and this can contribute to the total dissipative work differently. As we have discussed, base-pair-shearing events are seen for the OS3 and the OS5 pulling protocols, whereas base-pair-tearing events are observed for the EU and the MU protocols. Evidently, $\langle W \rangle$ is larger for the former two protocols than the latter two protocols. Therefore, friction or dissipation is much less for the tearing than the shearing of the DNA base pairs.

Note that the pulling force $F$ and hence $\langle W \rangle$ for the dissociation of dsDNA also depend on the pulling velocity $V$.[69] To check whether the conclusions remain unaltered by changing $V$, we compare $\langle W \rangle$–extension behavior for the B-DNA and the DNA–daunomycin complex when pulled via the SE stretching [Fig. 9(a)] and the MU [Fig. 9(b)] protocols for two different pulling velocities. For each of the two pulling protocols, we find that the absolute value of the work done for a given complex at a given DNA-extension depends very much on the pulling velocity $V$ and it decreases by lowering $V$. The magnitude of this decrease in the work done by lowering the pulling velocity also depends on the pulling protocol. Interestingly, we find almost similar $\langle W \rangle$–extension profile for the DNA–daunomycin complex for the SE pulling protocol, even by lowering $V$ from 0.2 Å/ns to 0.05 Å/ns [see Fig. 9(a)]. This means that for the SE pulling protocol, the deformation becomes reversible for $V \leq 0.05$ Å/ns. However, for the other pulling protocols, the critical pulling velocity $V_c$ after which the deformation becomes reversible might vary, as the response of $\langle W \rangle$ on $V$ is nonlinear and strongly depends on the pulling protocol. Importantly, we find from Fig. 9 that $\langle W \rangle$ for the DNA–daunomycin complex is always higher than that for the bare B-DNA for both the pulling protocols, at each pulling velocity. Therefore, our claim that DNA intercalator enhances the mechanical strength of dsDNA is robust, considering the five different pulling protocols, the multiple pulling velocities and the two intercalators studied in this work.

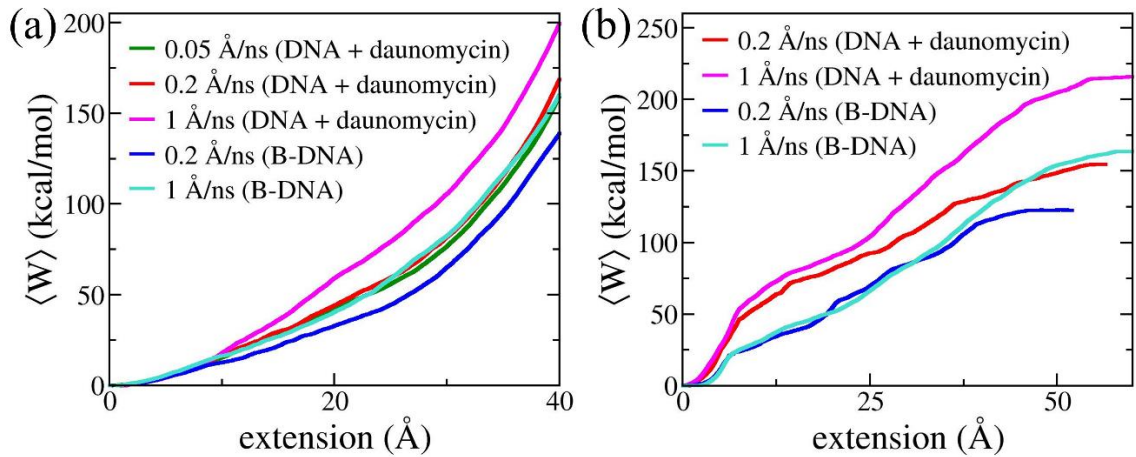

FIG. 9. Pulling velocity dependence of the average cumulative work done, $\langle W \rangle$, for the SE pulling (a) and the MU (b) protocols. The SE pulling and the MU protocols are described in Figs. 2(a) and 2(e), respectively. In each plot, $\langle W \rangle$ profiles at two different pulling velocities ($V = 0.2$ Å/ns and 1 Å/ns) are provided for both the DNA–daunomycin complex and the bare B-DNA. In (a), $\langle W \rangle$ profile for the DNA–daunomycin complex at a very small $V$ (= 0.05 Å/ns) is also plotted. Interestingly, this $\langle W \rangle$ profile is almost like that at $V = 0.2$ Å/ns.

## V. CONCLUSIONS

Using both equilibrium MD and SMD simulations, we show that DNA intercalator strongly modifies the mechanical properties of DNA. This might have serious consequences for active biological processes, such as DNA repair, transcription and replication. Particularly, we find that the stretch modulus, $\gamma$, of the DNA significantly increases upon intercalation of daunomycin or ethidium. The observed increase in $\gamma$ upon drug intercalation is in accord with recent experimental studies.[23,24] In contrast, we find that the persistence length $l_p$ and the bending modulus $\kappa$ of dsDNA decrease upon intercalation of daunomycin or ethidium. There results for the change in $l_p$ of DNA upon drug intercalations reported in various experimental studies disagree with each other, as discussed in Ref. 13. So, there is a scope for simulation studies to understand this discrepancy in the measured elastic moduli of different intercalated drug–DNA complexes.[24-26]

From all the pulling protocols, we find that dsDNA stretching and melting (dissociation of the two DNA strands) in the presence of either one of the intercalators become difficult because of the enhanced stability of intra-base-pair hydrogen bonds that arises from the π-stacking interaction between the phenyl rings of the DNA bases and of the intercalator. We also observe that the effect of DNA intercalators on the structure of DNA is local. Overall, our study is a significant step towards a quantitative understanding of the nanomechanics of DNA–ligand complexes and will be valuable for the DNA-targeted therapeutics research.[5] For designing better drugs for chemotherapy, our study calls for further in vivo investigations to understand the relation of the enhanced mechanical strength of DNA in presence of intercalated anticancer drug found in this study to its biological functions.

It would be interesting to study the thermal denaturation[70] of dsDNA in the presence and absence of intercalators and compare the temperature-induced melting pathways to the force-induced melting pathways obtained in this study by the various pulling protocols. Given that people have already started to use DNA as molecular wire in nanoelectronics, how intercalators modify DNA's electrical properties[71,72] is of technological and fundamental interest. Some of these aspects are currently under investigation in our group.

## SUPPLEMENTARY MATERIAL

See supplementary material for figures S1–4. Rise and twist versus the base pair step for all the three complexes, analyses of the structural and elastic properties of the middle 6 base pairs of DNA for all the three complexes, and the instantaneous snapshots of the DNA–ethidium complex at various DNA-extensions for the OS3 pulling and the MU protocols are given.

## ACKNOWLEDGMENTS

We thank the Supercomputer Education and Research Centre (SERC), IISc, for providing the computational facility at SAHASRAT machine. A.K.S. thanks MHRD for the research fellowship.

# Supplementary Material

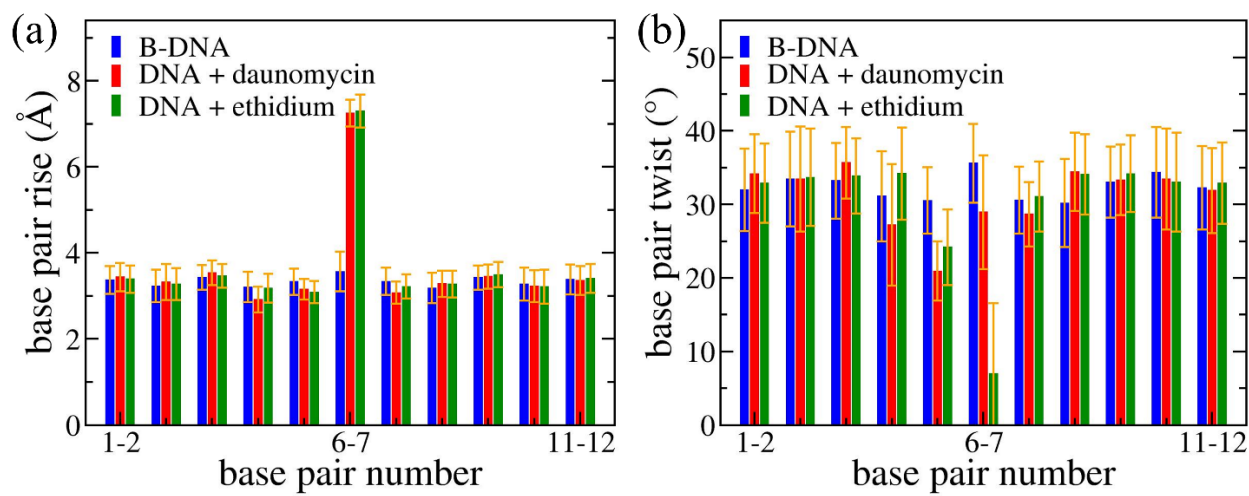

FIG. S1. Effect of an intercalator on DNA's inter base pair rise and twist. Rise (a) and twist (b) as a function of the base pairs step (number) are plotted for the bare B-DNA and the two intercalated drug–DNA complexes. The error bar for each step of base pairs in each plot is the standard deviation of 2500 data points obtained from the last 50 ns of simulation time. For each drug intercalated complex, rise for the step of the 6–7 base pairs is ~3.5 Å more than that of the bare B-DNA, as an intercalator is present between these two base pairs. Otherwise, the rise is the same both in the presence and absence of an intercalator. Twist for the step of the 6–7 base pairs of the DNA–daunomycin (= 29°) and DNA–ethidium (= 7°) complexes are much less than that of the B-DNA (= 36°). Twist of DNA in the presence of an intercalated drug differs from that of the bare B-DNA maximum up to 3 base pairs on either side of the intercalation site. This represents local unwinding of the DNA double helix structure in the presence of an intercalator, which is more significant for ethidium than daunomycin.

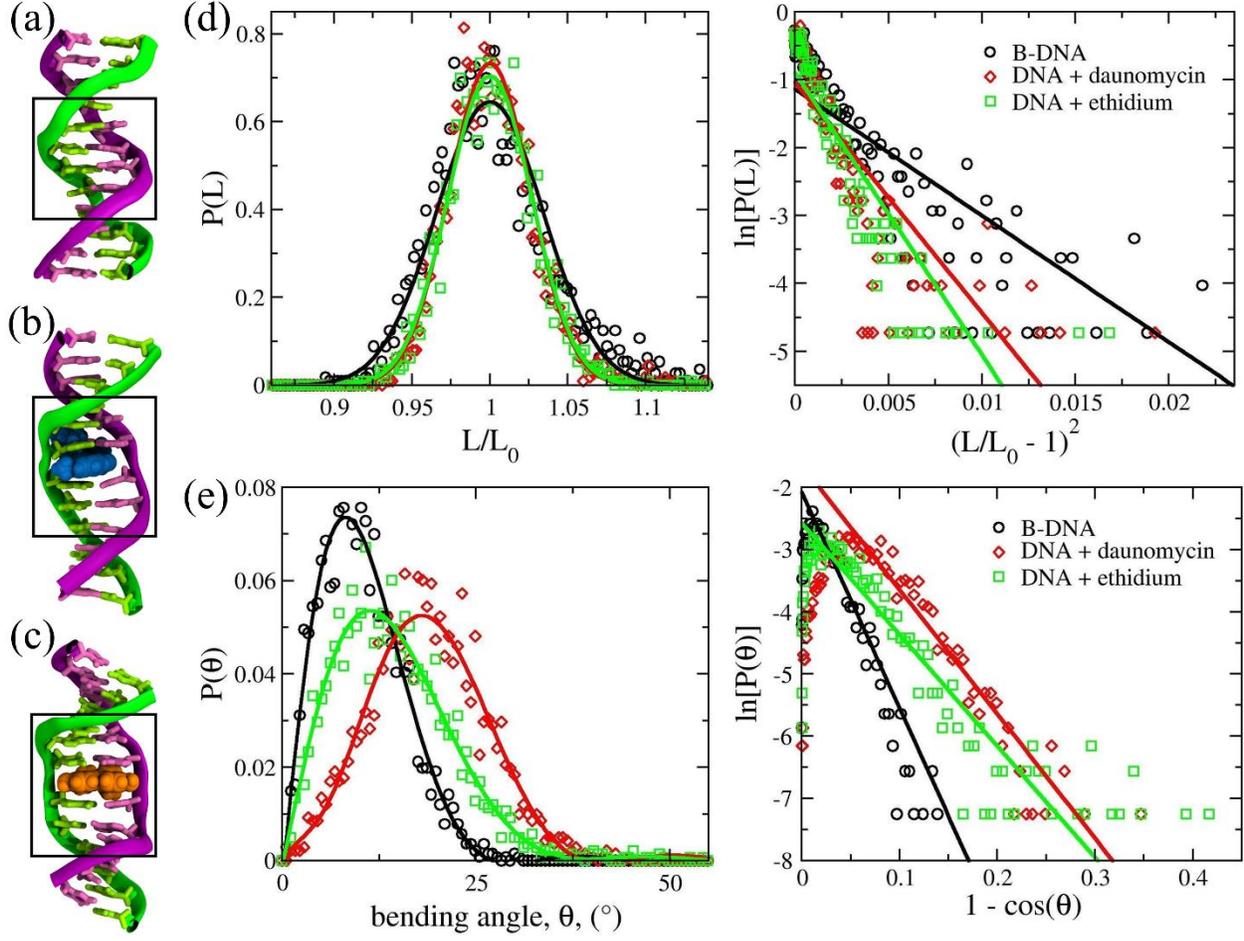

FIG. S2. Comparison of structure and elastic properties of the middle 6 base-pairs of DNA (marked within the rectangle) for the bare B-DNA (a), the DNA–daunomycin (b), and the DNA–ethidium (c) from equilibrium simulations. The representations and colors are the same as shown in Fig. 1 in the main text. Stretch modulus, $\gamma$, and persistence length, $l_P$, (as well as bending rigidity $\kappa$) for each system given in Table II in the main text are calculated from the normalized probability distributions of DNA contour length (d) and bending angle (e), respectively. See the calculation details in the main text. The distributions for each system are obtained from the last 50 ns of 200 ns long simulation.

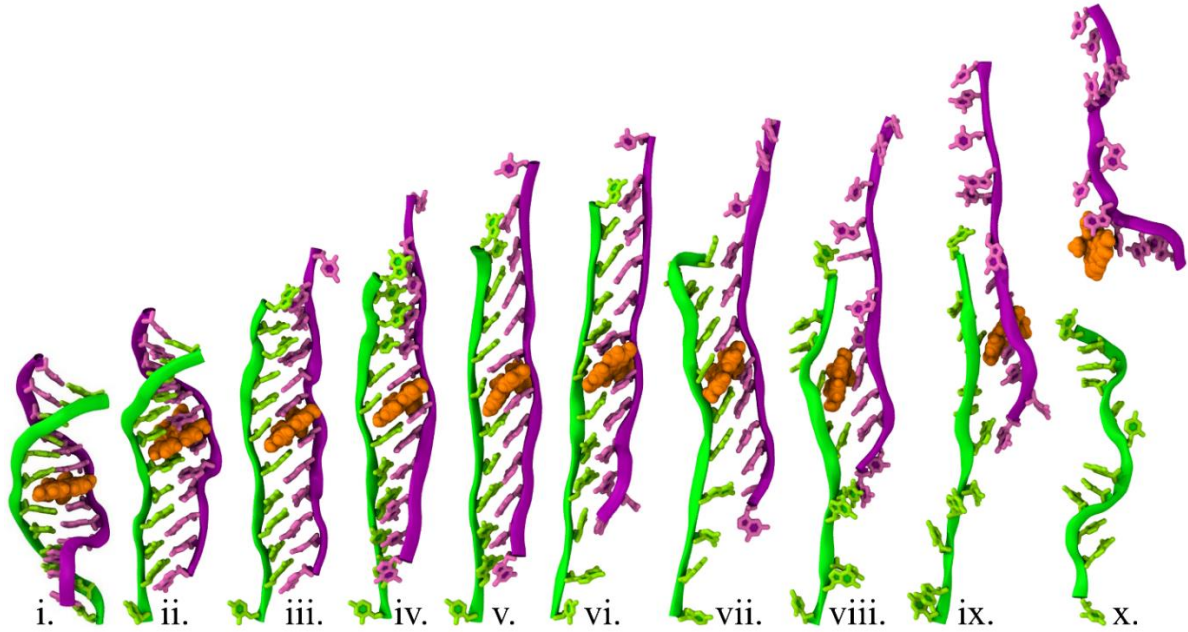

FIG. S3. (i–x) The instantaneous snapshots for shearing of the DNA–ethidium complex by the OS3 pulling protocol are shown at various DNA-extensions as pointed in the force–extension plot in Fig. 8(a) in the main text. The representation and color coding in each snapshot are the same as described in Fig. 1 in the main text.

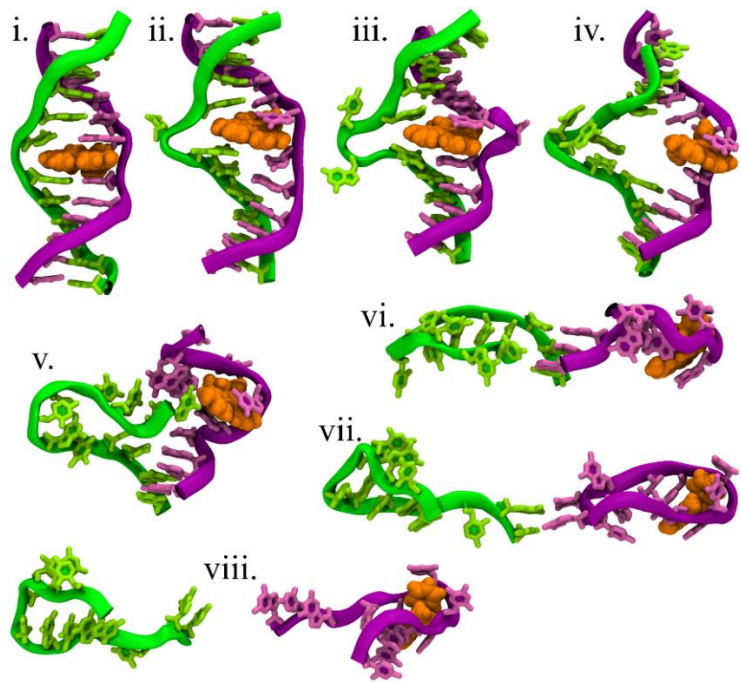

FIG. S4. (i–viii) The instantaneous snapshots for tearing of the DNA–ethidium complex by the MU protocol are shown at various DNA-extensions as pointed in the force–extension plot in Fig. 8(b) in the main text. The representation and color coding in each snapshot are the same as described in Fig. 1 in the main text.